\documentclass{webofc}
\wocname{European Physical Journal Web of Conferences}
\woctitle{International Workshop on Multi facets of Eos and Clustering (IWM-EC 2014)}
\begin{document}
\title{Symmetry energy systematics and its high density behavior}
%
%

\author{Lie-Wen Chen\inst{1}\fnsep\thanks{\email{lwchen@sjtu.edu.cn}} }
\institute{Department of Physics and Astronomy and Shanghai Key Laboratory for
Particle Physics and Cosmology, Shanghai Jiao Tong University, Shanghai 200240, China}

\abstract
{
We explore the systematics of the density dependence of nuclear matter
symmetry energy in the ambit of microscopic calculations with various energy density
functionals, and find that the symmetry energy from subsaturation density to
supra-saturation density can be well determined by three characteristic parameters of
the symmetry energy at saturation density $\rho_0 $, i.e., the magnitude
$E_{\text{sym}}({\rho_0 })$, the density slope $L$ and the density curvature
$K_{\text{sym}}$. This finding opens a new window to constrain the supra-saturation
density behavior of the symmetry energy from its (sub-)saturation density behavior. In particular, we obtain
$L=46.7 \pm 12.8$ MeV and $K_{\text{sym}}=-166.9 \pm 168.3$ MeV
as well as $E_{\text{sym}}({2\rho _{0}}) \approx 40.2 \pm 12.8$
MeV and $L({2\rho _{0}}) \approx 8.9 \pm 108.7$ MeV based on the present knowledge of
$E_{\text{sym}}({\rho_{0}}) = 32.5 \pm 0.5$ MeV, $E_{\text{sym}}({\rho_c}) = 26.65 \pm 0.2$ MeV 
and $L({\rho_c}) = 46.0 \pm 4.5$ MeV at $\rho_{\rm{c}}= 0.11$ fm$^{-3}$ extracted from 
nuclear mass and the neutron skin thickness of Sn isotopes. Our results indicate that the
symmetry energy cannot be stiffer than a linear density dependence. 
In addition, we also discuss the quark matter symmetry energy since the deconfined 
quarks could be the right degree of freedom in dense matter at high baryon densities.
}
\maketitle

\section{Introduction}

The nuclear matter symmetry energy, which essentially characterizes the
isospin dependent part of the equation of state (EOS) of asymmetric
nuclear matter, is important for understanding many questions in 
nuclear physics and astrophysics, including the nuclear
effective interactions in asymmetric nuclear matter, the structure
and stability of exotic nuclei, the reaction dynamics induced by rare
isotopes, the nature and evolution of neutron stars, and the mechanism 
of supernova explosion ~\cite%
{LiBA98,Dan02a,Lat04,Ste05a,Bar05,CKLY07,LCK08}. The symmetry energy 
also plays an important role in some interesting issues of new physics
beyond the standard model~\cite{Hor01b,Sil05,Wen09,Zhe14,Zhe15}. During the 
last decade, a lot of experimental, observational and theoretical efforts 
have been devoted to constraining the density dependence of the symmetry 
energy~\cite{LiBAEPJA14,Tsa12,Lat12,ChenLW12,LiBA12,Hor14}. While significant progress
has been made in determining the density behavior of the symmetry energy
around saturation density ${\rho _{0}}$ ($\sim 0.16$ fm$^{-3}$), its
supra-saturation density behavior is still poorly known and remains the
most uncertain property of isospin asymmetric nuclear matter. Theoretically, 
many experimental and observational probes have been proposed to extract 
information on the supra-saturation density behavior of the symmetry 
energy~\cite{LiBAEPJA14}. In terrestrial laboratories, heavy-ion collisions 
provide the only way to explore the supra-saturation density behavior
of the symmetry energy under controlled conditions. 
To the best of our knowledge, the constraints on the supra-saturation density 
behavior of the symmetry energy obtained so far are all from the transport model
analyses on the data of $\pi^- / \pi^+$ ratio~\cite{Xiao09,Fen10,XuJ10,Xie13,XuJ13,Hon14,Son14} 
and $n/p$ elliptic flows~\cite{Rus11,Coz13} in heavy-ion collisions but 
unfortunately they are contradictive with each other, leaving a confusing 
situation for the community.

Conventionally, the nuclear matter EOS is defined as the binding energy 
per nucleon as a function of the density and a number of bulk characteristic 
parameters defined at saturation density $\rho_0 $ are usually introduced to 
quantitatively characterize the energy of symmetric nuclear matter and 
the symmetry energy~\cite{Che09,Che11a}. For example, the energy $E_0(\rho_0)$ and
incompressibility $K_0$ of symmetric nuclear matter are the two
lowest order bulk parameters for the EOS of symmetric nuclear matter
while the symmetry energy magnitude $E_{\text{sym}}(\rho_0)$ and its slope parameter
$L $ are the two lowest order bulk parameters of 
the symmetry energy. 
While several lower order bulk characteristic parameters of asymmetric 
nuclear matter, such as $E_0(\rho_0)$, $K_0$, $E_{\text{sym}}(\rho_0)$ and 
$L$ have been relatively well constrained or in significant 
progress~\cite{LiBAEPJA14,Tsa12,Lat12,ChenLW12,LiBA12,Hor14,You99}, yet 
the higher order bulk characteristic parameters are still poorly known. 
Actually, there has little experimental information on the
third-order derivative parameter $J_0$ of symmetric nuclear matter
at $\rho_0 $~\cite{Cai14} and the symmetry energy curvature parameter 
$K_{\text{sym}}$~\cite{Che11a}. However, the higher order bulk characteristic parameters have been
shown to be closely related to some important issues in nuclear
physics and astrophysics, such as the determination of the isobaric
incompressibility of asymmetric nuclear matter~\cite{Che09,Che09a}
and the core-crust transition density and pressure in neutron
stars~\cite{Xu09a,Xu09b,Duc10}. In particular, within the Skyrme 
energy density functional, it has been proposed~\cite{Che11a} that the 
higher-order curvature parameter $K_{\text{sym}}$ may play an important role 
in the determination of the supra-saturation density behaviors of the 
symmetry energy.

So far (very likely also in future), essentially all the obtained 
constraints on $E_{\text{\textrm{sym}}}(\rho )$ are based on some energy
density functionals or phenomenological parameterizations of
$E_{\text{\textrm{sym}}}(\rho )$. Therefore, it would be very 
interesting to see whether there exist some universal 
laws (systematics) for the density dependence of the symmetry energy
within these functionals or parameterizations and whether one can
get some useful information on the high density symmetry energy from the 
relatively well-known knowledge of the symmetry energy around saturation density. 
For all the energy density functionals or phenomenological parameterizations, the
$E_{\text{sym}}(\rho )$ increases from $\rho = 0$ up to a certain
density around ${\rho _{0}}$ and then either continuously increases or
decreases depending on the parameters of the energy density functionals or
phenomenological parameterizations. While the parameters 
$E_{\text{sym}}({\rho_0 })$, $L$ and $K_{\text{sym}}$ accurately
characterize the symmetry energy density behaviors around $\rho_0 $, their
relation to the density behaviors at sub- and supra-saturation densities
in various energy density functionals or phenomenological parameterizations 
of $E_{\text{\textrm{sym}}}(\rho )$ are still unclear.

In the present talk, we report the preliminary results of the study on the
systematics of the density dependence of nuclear matter symmetry energy in 
the ambit of microscopic calculations with various energy density functionals. 
We systematically analyze the relation between the parameters
$E_{\text{sym}}({\rho_0 })$, $L$ and $K_{\text{sym}}$ defined at saturation 
density $\rho_0$ and the symmetry energy density behaviors at sub- and 
supra-saturation densities in various energy density functionals. 
In addition, since the dense matter at high baryon densities could be quark 
matter, we also discuss briefly the quark matter symmetry energy.

\section{Symmetry energy systematics}

The EOS of isospin asymmetric nuclear matter, given by its binding energy
per nucleon, can be expanded to $2$nd-order in isospin asymmetry $\delta $ as%
\begin{equation}
E(\rho ,\delta )= E_{0}(\rho )+E_{\mathrm{sym}}(\rho )\delta ^{2}+O(\delta
^{4}),  \label{EOSANM}
\end{equation}%
where $\rho =\rho _{n}+\rho _{p}$ is the baryon density with $\rho _{n}$ and
$\rho _{p}$ denoting the neutron and proton densities, respectively; $\delta
=(\rho _{n}-\rho _{p})/\rho $ is the isospin asymmetry; $E_{0}(\rho )=E(\rho
,\delta =0)$ is the binding energy per nucleon in symmetric nuclear matter,
and the nuclear symmetry energy is expressed as%
\begin{equation}
E_{\mathrm{sym}}(\rho ) =\frac{1}{2!}\frac{\partial ^{2}E(\rho ,\delta )}{%
\partial \delta ^{2}}|_{\delta =0}.
\end{equation}%
Around a reference density $\rho _{r}$, the $E_{\mathrm{sym}}(\rho )$ can
be expanded in $\chi_r=(\rho -{\rho _{r}})/\rho _{r}$ as
\begin{equation}
E_{\text{sym}}(\rho )=E_{\text{sym}}({\rho _{r}})+\frac{L(\rho _{r})}{3}\chi_r+\frac{K_{\mathrm{sym}}(\rho _{r})}{2!}\chi_r
^{2}+O(\chi_r ^{3}),
\end{equation}
where $L(\rho _{r})=3{\rho _{r}}\frac{\partial E_{\mathrm{sym}}(\rho )}{\partial
\rho }|_{\rho ={\rho _{r}}}$ and $K_{%
\mathrm{sym}}(\rho _{r})=9\rho _{r}^{2}\frac{d^{2}E_{\mathrm{sym}}(\rho )}{d\rho ^{2}}%
|_{\rho =\rho _{r}}$ are, respectively, the slope and curvature parameters of the 
symmetry energy at $\rho _{r}$, and they are the lowest-order two bulk 
parameters characterizing the density behaviors of the symmetry energy 
around $\rho _{r}$. In particular, when the reference density $\rho _{r}$ is taken 
as the saturation density $\rho _{0}$, the $L(\rho _{r})$ and $K_{\mathrm{sym}}(\rho _{r})$ 
are then reduced to the famous symmetry energy slope parameter $L\equiv 3\rho
_{0}\frac{dE_{\mathrm{sym}}(\rho )}{d\rho }|_{\rho =\rho _{0}}$ and symmetry energy 
curvature parameter $K_{%
\mathrm{sym}}\equiv 9\rho _{0}^{2}\frac{d^{2}E_{\mathrm{sym}}(\rho )}{d\rho ^{2}}%
|_{\rho =\rho _{0}}$, respectively.

\begin{figure}[tbp]
\centering
\includegraphics[scale=0.3]{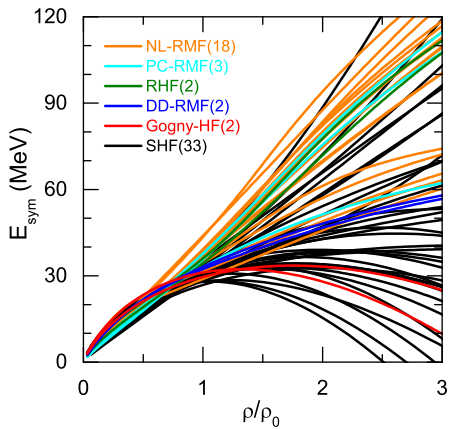}
\caption{(Color online) The symmetry energy as a function of the density
normalized by the corresponding saturation $\rho_0$ in various energy
density functionals with $60$ interactions. See text for the details.}
\label{EsymRho}
\end{figure}

To examine the symmetry energy systematics, we select a comprehensive
large sample of $60$ well-calibrated interactions in various energy density
functionals, namely, $33$ Skyrme interactions (v090, MSk7, BSk8, SKP, SKT6,
SKX, BSk17, SGII, SKM*, SLy4, SLy5, MSkA, MSL0, SIV, SkSM*, kMP, SKa, Rsigma,
Gsigma, SKT4, SV, SkI2, SkI5, BSK18, BSK19, BSK20, BSK21, MSL1, SAMi, SV-min,
UNEDF0, UNEDF1, TOV-min), $2$ Gogny interactions (D1S and D1N), $18$ nonlinear
RMF interactions (FSUGold, PK1s24, NL3s25, G2, TM1, NL-SV2, NL-SH, NL-RA1, PK1,
NL3, NL3*, G1, NL2, NL1, IU-FSU, BSP, IUFSU*, TM1*), $2$ density-dependent RMF
interactions (DD-ME1 and DD-ME2), $3$ point-coupling RMF interactions (DD-PC1,
PC-PK1, PC-F1), and $2$ relativistic HF interactions (PKO3 and PKA1). These
interactions include the $46$ interactions used in Ref.~\cite{Maz11}
(except BCP which is designed for density up to only $0.24$ fm$^{-3}$)
and other $14$ interactions (i.e., BSK18, BSK19, BSK20, BSK21, MSL1, SAMi, SV-min,
UNEDF0, UNEDF1, TOV-min, IU-FSU, BSP, IUFSU*, TM1*) constructed more recently.
Shown in Fig.~\ref{EsymRho} is the symmetry energy as a function
of the density normalized by the corresponding saturation $\rho_0$ with the $60$
interactions. It is clearly seen that various energy density functionals predict 
very different density behaviors of the symmetry energy, especially at supra-saturation 
densities. For example, the magnitude of the symmetry energy at $2\rho_0$ can be 
varied from about $15$ MeV to $100$ MeV, depending on the models and interaction 
parameters. Furthermore, it is seen that some non-relativistic interactions
predict negative symmetry energy at baryon densities above about $2.5\rho_0$.

\begin{figure}[tbp]
\centering
\includegraphics[scale=0.38]{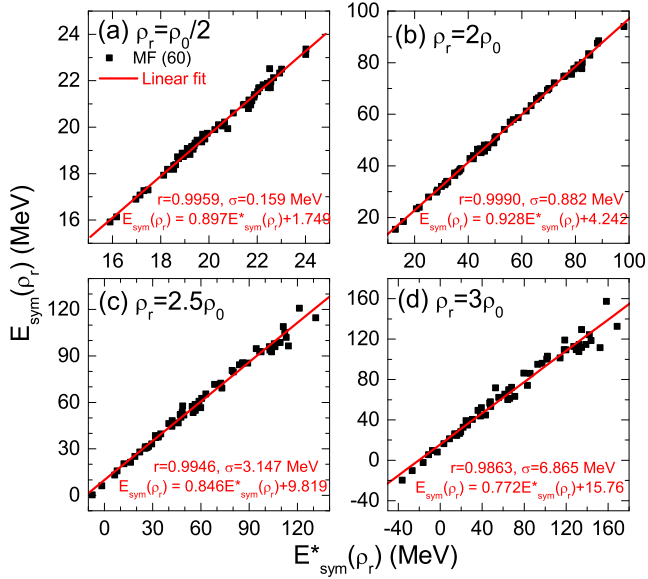}
\caption{(Color online) $E_{\mathrm{sym}}(\rho )$
vs $E^{*}_{\mathrm{sym}}(\rho )$ in various energy
density functionals with $60$ interactions at
$\rho = 0.5\rho_0$, $\rho = 2\rho_0$, $\rho = 2.5\rho_0$ and
$\rho = 3\rho_0$.}
\label{EsymCorr}
\end{figure}

\begin{figure}[tbp]
\centering
\includegraphics[scale=0.38]{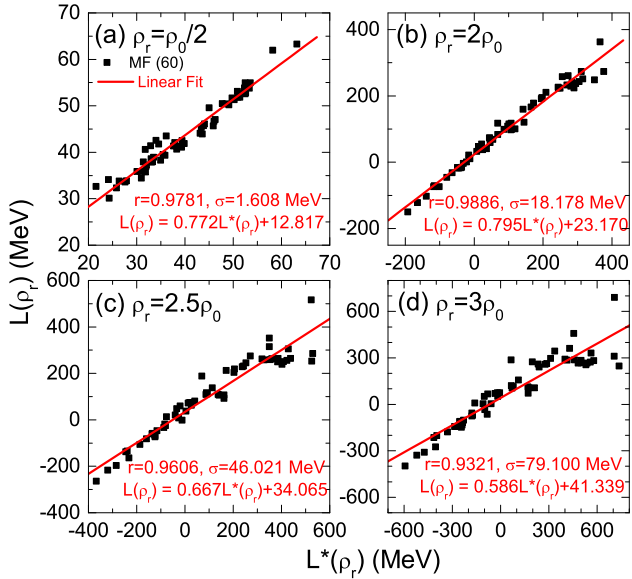}
\caption{(Color online) Same as Fig.~\ref{EsymCorr} but for $L(\rho )$
vs $L^{*}(\rho )$.}
\label{LCorr}
\end{figure}

Around the saturation density $\rho_0$, the magnitude $E_{\mathrm{sym}}(\rho )$ 
and the density slope $L(\rho )$ of the symmetry energy can be well approximated, 
respectively, by
\begin{equation}
E^{*}_{\mathrm{sym}}(\rho ) \equiv E_{\mathrm{sym}}(\rho _{0})+L\chi +\frac{K_{\mathrm{sym}}}{2!}\chi
^{2},
\label{Esymst}
\end{equation}%
and
\begin{equation}
L^{*}(\rho ) \equiv L\rho/\rho_0 + K_{\mathrm{sym}}\chi \rho/\rho_0,
\label{Lst}
\end{equation}%
with $\chi=(\rho -{\rho _{0}})/\rho _{0}$. Based on a correlation analysis 
within SHF approach, it has been shown~\cite{Che11a} that $E^{*}_{\mathrm{sym}}(\rho )$ 
can well describe the magnitude of the symmetry energy up to $2\rho_0$. 
How well can Eq.~(\ref{Esymst}) and Eq.~(\ref{Lst}) approximate the 
corresponding values at densities deviated from $\rho_0$ in various energy 
density functionals? Shown in Fig.~\ref{EsymCorr} is $E_{\mathrm{sym}}(\rho )$ 
vs $E^{*}_{\mathrm{sym}}(\rho )$ with the $60$ interactions at
$\rho = 0.5\rho_0$, $\rho = 2\rho_0$, $\rho = 2.5\rho_0$ and
$\rho = 3\rho_0$. A very strong linear correlation (the Pearson linear 
correlation coefficient $r$ is larger than $0.98$ for all the cases) is 
observed between $E_{\mathrm{sym}}(\rho )$ and $E^{*}_{\mathrm{sym}}(\rho )$
for the $60$ interactions at all the four densities considered here.
Similarly, Fig.~\ref{LCorr} shows the correlation between $L(\rho )$ and 
$L^{*}(\rho )$ with the $60$ interactions at $\rho = 0.5\rho_0$, 
$\rho = 2\rho_0$, $\rho = 2.5\rho_0$ and $\rho = 3\rho_0$, and again a strong 
linear correlation is observed between $L(\rho )$ and $L^{*}(\rho )$ 
($r$ is larger than $0.93$ for the densities considered here).

The strong linear correlation between $E_{\mathrm{sym}}(\rho )$
and $E^{*}_{\mathrm{sym}}(\rho )$ as well as between $L(\rho )$ and
$L^{*}(\rho )$ for the $60$ interactions at different densities shown
in Fig.~\ref{EsymCorr} and Fig.~\ref{LCorr} leads to the following
relations
\begin{eqnarray}
E_{\mathrm{sym}}(\rho ) &\approx & A(\rho ) + B(\rho )E^{*}_{\mathrm{sym}}(\rho ), \label{EsymSysEq}\\
L(\rho ) &\approx & A_{\text L}(\rho ) + B_{\text L}(\rho )L^{*}(\rho ).
\label{LSysEq}
\end{eqnarray}%
The values of the coefficients $A(\rho )$, $B(\rho )$, $A_{\text L}(\rho )$ and 
$B_{\text L}(\rho )$ generally depend the density. In particular, one can see 
from Fig.~\ref{EsymCorr} and Fig.~\ref{LCorr} that $A(\rho )$ ($A_{\text L}(\rho )$) 
is generally nonzero and $B(\rho )$ ($B_{\text L}(\rho )$) usually deviates from 
unit, reflecting the higher-order effects beyond the expansion in Eq.~(\ref{Esymst}) 
and Eq.~(\ref{Lst}). We note $A(\rho )\approx 0$ ($A_{\text L}(\rho )\approx 0$) 
and $B(\rho )\approx 1$ ($B_{\text L}(\rho )\approx 1$) for $\rho \approx \rho_0$ 
as expected. Eq.~(\ref{EsymSysEq}) and Eq.~(\ref{LSysEq}) establish the systematics 
of $E_{\mathrm{sym}}(\rho )$ and $L(\rho )$ in terms of three characteristic
parameters $E_{\text{sym}}({\rho_0 })$, $L$ and $K_{\text{sym}}$. It should be 
noted that in principe Eq.~(\ref{LSysEq}) can also be deduced from Eq.~(\ref{EsymSysEq}) 
according to the definition. We would like 
to point out the Eq.~(\ref{EsymSysEq}) for the systematics of $E_{\mathrm{sym}}(\rho )$ 
can be safely applied in the density region from $\rho_0 /5$ to $3\rho_0$ where the 
Pearson linear correlation coefficient $r$ is always larger than $0.96$. Similarly, 
the Eq.~(\ref{LSysEq}) for the systematics of $L(\rho )$ can be safely applied in 
the density region from $\rho_0 /2$ to $2.5\rho_0$ where the
Pearson linear correlation coefficient $r$ is always larger than $0.96$.

\section{Supra-saturation density behaviors of the symmetry energy}

The systematics of $E_{\mathrm{sym}}(\rho )$ and $L(\rho )$ in Eq.~(\ref{EsymSysEq}) 
and Eq.~(\ref{LSysEq}) imply that the three characteristic parameters 
$E_{\text{sym}}({\rho_0 })$, $L$ and $K_{\text{sym}}$ (and thus 
$E_{\mathrm{sym}}(\rho )$ and $L(\rho )$) can be determined once 
three values of either $E_{\mathrm{sym}}(\rho )$ or $L(\rho )$ are 
known. This means that one can extract information on the high density behaviors 
of the symmetry energy from the relatively well constrained (sub-)saturation 
density behaviors of the symmetry energy.

\begin{figure}[tbp]
\centering
\includegraphics[scale=0.32]{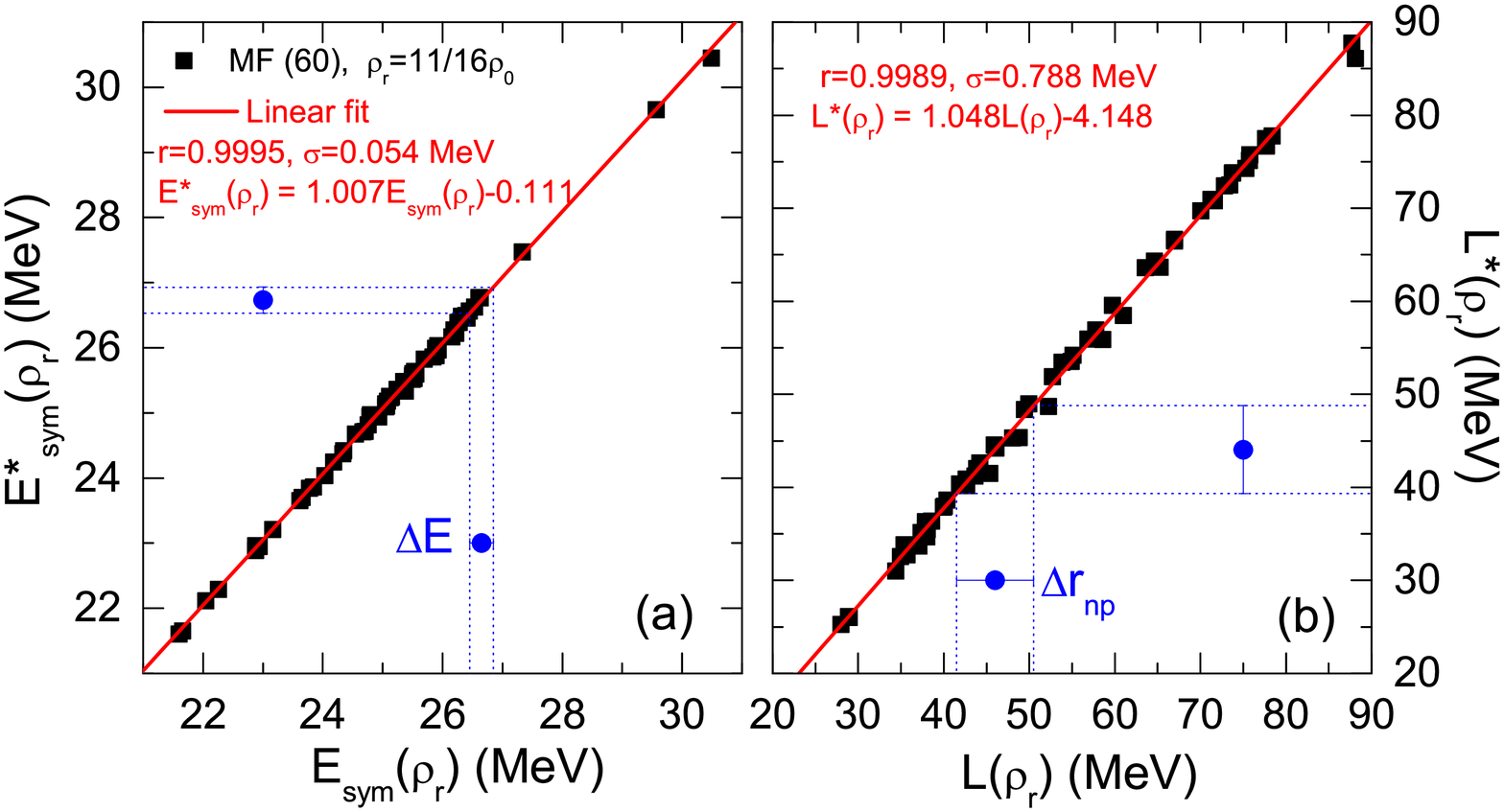}
\caption{(Color online) $E_{\text{sym}}(\rho_c )$ vs $E^{*}_{\text{sym}}(\rho_c )$ (a)
and $L(\rho_c )$ vs $L^{*}(\rho_c )$ (b) in various energy
density functionals with the $60$ interactions. The constraints 
$E_{\text{sym}}(\rho_{\text{c}})=26.65 \pm 0.20$ MeV ($\Delta E$) and 
$L(\rho_{\text{c}})=46.0 \pm 4.5$ MeV ($\Delta r_{np}$) obtained 
in Ref.~\cite{Zha13} are also included.}
\label{EsymLrhoC}
\end{figure}

In recent years, several accurate constraints on the symmetry energy at 
subsaturation density have been obtained through analyzing nuclear 
structure properties of heavy nuclei. Indeed, a quite accurate constraint 
on the symmetry energy at the subsaturation cross density 
$\rho_{\text{c}} = 0.11$ fm$^{-3}$, i.e., 
$E_{\text{sym}}(\rho_{\text{c}})=26.65 \pm 0.20$ MeV, has been recently
obtained from analyzing the binding energy difference of heavy isotope
pairs~\cite{Zha13}. At the same time, an accurate constraint on the 
density slope at $\rho_{\text{c}}$, i.e, $L(\rho_{\text{c}})=46.0 \pm 4.5$ MeV 
has been obtained from analyzing the neutron skin data of Sn 
isotopes~\cite{Zha13}. At density $\rho_{\text{c}} = 0.11$ fm$^{-3}$, 
the systematics of $E_{\mathrm{sym}}(\rho )$ and $L(\rho )$ in Eq.~(\ref{EsymSysEq})
and Eq.~(\ref{LSysEq}) lead to the following expressions
\begin{eqnarray}
E^{*}_{\text{sym}}(\rho_c ) &\approx & a(\rho_c ) + b(\rho_c )E_{\text{sym}}(\rho_{\text{c}}),\\
L^{*}(\rho_c ) &\approx & a_{\text L}(\rho_c ) + b_{\text L}(\rho_c )L(\rho_c ).
\end{eqnarray}%
The values of coefficients $a(\rho_c )$ and $b(\rho_c )$ ($a_{\text L}(\rho_c )$ 
and $b_{\text L}(\rho_c )$) can be obtained from linear fitting to the correlation 
between $E^{*}_{\text{sym}}(\rho_c )$ and $E_{\text{sym}}(\rho_{\text{c}})$
($L^{*}(\rho_c )$ and $L(\rho_c )$). Shown in Fig.~\ref{EsymLrhoC} are 
$E_{\text{sym}}(\rho_c )$ vs $E^{*}_{\text{sym}}(\rho_c )$ and $L(\rho_c )$ vs 
$L^{*}(\rho_c )$ with the $60$ interactions, and one can observe a very strong linear 
correlation (the Pearson linear correlation coefficient $r$ is about $0.999$ for 
both cases) between $E^{*}_{\text{sym}}(\rho_c )$ and $E_{\text{sym}}(\rho_{\text{c}})$
as well as between $L^{*}(\rho_c )$ and $L(\rho_c )$, and these linear correlations 
lead to $a(\rho_c ) = -0.111 \pm 0.111$ MeV, $b(\rho_c ) = 1.007 \pm 0.004$, 
$a_{\text L}(\rho_c ) = -4.148 \pm 0.358$ MeV and $b_{\text L}(\rho_c ) = 1.048 \pm 0.006$.

Besides $E_{\text{sym}}(\rho_{\text{c}})$ and $L(\rho_{\text{c}})$, one needs 
another constraint condition to determine $E_{\text{sym}}({\rho_0 })$, $L$ and 
$K_{\text{sym}}$ and thus the supra-saturation density behaviors of the 
symmetry energy. In the present work, we further use the constraint of 
$E_{\text{sym}}({\rho _{0}}) = 32.5\pm0.5$ MeV obtained recently from a new 
and more accurate finite-range droplet model analysis of the nuclear
mass of the 2003 Atomic Mass Evaluation~\cite{Mol12}. Therefore, from 
$E_{\text{sym}}(\rho_{\text{c}})=26.65 \pm 0.20$ MeV, 
$L(\rho_{\text{c}})=46.0 \pm 4.5$ MeV and 
$E_{\text{sym}}({\rho _{0}}) = 32.5\pm0.5$ MeV, one can obtain 
$L=46.7 \pm 12.8$ MeV and $K_{\text{sym}}=-166.9 \pm 168.3$ MeV.
It is interesting to see that the obtained $L=46.7 \pm 12.8$ MeV is in very 
good agreement with other constraints extracted from terrestrial experiments, 
astrophysical observations, and theoretical calculations with controlled
uncertainties~\cite{LiBAEPJA14,Tsa12,Lat12,ChenLW12,LiBA12,Hor14}. The obtained 
$K_{\text{sym}}=-166.9 \pm 168.3$ MeV also agrees well with the result 
$K_{\text{sym}}=-100 \pm 165$ MeV~\cite{Che11a} obtained from a correlation analysis
within SHF approach.

\begin{figure}[tbp]
\centering
\includegraphics[scale=0.35]{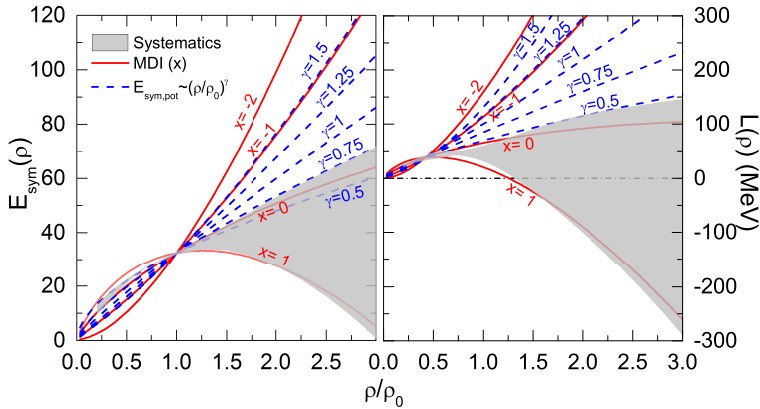}
\caption{(Color online) 
Density dependence of $E_{\mathrm{sym}}(\rho )$ and $L(\rho )$ according to 
the systematics in Eq.~(\ref{EsymSysEq}) and Eq.~(\ref{LSysEq}) with 
$E_{\text{sym}}({\rho _{0}}) = 32.5\pm0.5$ MeV,
$L=46.7 \pm 12.8$ MeV and $K_{\text{sym}}=-166.9 \pm 168.3$ MeV.
The results from the MDI interaction and the phenomenological 
parameterizations of $E_{\text{sym,pot}}(\rho ) \sim (\rho / \rho_0)^{\gamma}$ 
are also include for comparison. }
\label{EsymLRho}
\end{figure}

Based on $E_{\text{sym}}({\rho _{0}}) = 32.5\pm0.5$ MeV,
$L=46.7 \pm 12.8$ MeV and $K_{\text{sym}}=-166.9 \pm 168.3$ MeV, one 
then can obtain $E_{\mathrm{sym}}(\rho )$ and $L(\rho )$ according to 
the systematics in Eq.~(\ref{EsymSysEq}) and Eq.~(\ref{LSysEq}), and 
the results are shown in Fig.~\ref{EsymLRho}. For comparison, we also 
include the results from the MDI interaction~\cite{Che05} with 
$x = 1$, $0$, $-1$ and $-2$ and the phenomenological parameterizations of 
$E_{\text{sym,pot}}(\rho ) \sim (\rho / \rho_0)^{\gamma}$~\cite{Tsa09,Rus11} 
for the potential energy part of the symmetry energy with 
$\gamma = 0.5$, $0.75$, $1.0$, $1.25$ and $1.5$. We would like to point 
out that the MDI interaction and the parameterizations of 
$E_{\text{sym,pot}}(\rho ) \sim (\rho / \rho_0)^{\gamma}$ 
have been extensively applied in transport model simulations of 
heavy ion collisions. One can see from Fig.~\ref{EsymLRho} that the present 
analysis based on the symmetry energy systematics with 
$E_{\text{sym}}({\rho _{0}}) = 32.5\pm0.5$ MeV,
$L=46.7 \pm 12.8$ MeV and $K_{\text{sym}}=-166.9 \pm 168.3$ MeV favors a softer 
symmetry energy and suggests that the symmetry energy cannot be stiffer 
than a linear density dependence. 

In particular, at the supra-saturation density of $2\rho_0$, we find
$E_{\text{sym}}({2\rho _{0}}) = 40.2\pm12.8$ MeV and
$L({2\rho _{0}})=8.9 \pm 108.7$ MeV. We note that these values are in nice 
agreement with the variational many-body theory calculation with WFF1 
interaction~\cite{Wir88} which can give a good description on the recent observation of 
heavy neutron stars with radius of $9.1^{+1.3}_{-1.5}$ km~\cite{Gui13}.

\section{Quark matter symmetry energy}

At extremely high baryon density, the main degree of freedom could be the 
deconfined quark matter rather than the confined baryon matter, and there the 
quark matter symmetry energy should be involved for the properties of isospin 
asymmetric quark matter (isospin symmetry is still satisfied in quark matter). 
The isospin asymmetric quark matter could be produced in ultra-relativistic 
heavy ion collisions induced by neutron-rich nuclei and it could also exist 
in compact stars such as neutron stars or quark stars.
Although significant progress has been made in understanding the density 
dependence of the nuclear matter symmetry energy, there has little information 
on the density dependence of the quark matter symmetry energy. Theoretically,
it is difficult to calculate the quark matter symmetry energy since the 
{\it ab initio} Lattice QCD simulations does not work at finite baryon density 
while perturbative QCD only works at extremely high baryon density.

Similarly as in the case of nuclear matter, the EOS of quark matter consisting 
of $u$, $d$, and $s$ quarks, defined by its binding energy per baryon number, 
can be expanded in isospin asymmetry $\delta_q $ as
\begin{equation}
E(n_B ,\delta, n_s)=E_{0}(n_B, n_s)+E_{\mathrm{sym}}(n_B, n_s)\delta_q ^{2}+\mathcal{O}(\delta_q ^{4}),
\label{EOSAQM}
\end{equation}%
where $E_{0}(n_B, n_s)=E(n_B ,\delta_q =0, n_s)$ is the binding energy per
baryon number in three-flavor $u$-$d$-$s$ quark matter with equal fraction
of $u$ and $d$ quarks; the quark matter symmetry energy $E_{\mathrm{sym}}(n_B, n_s)$ is
expressed as
\begin{equation}
E_{\mathrm{sym}}(n_B, n_s) =\left. \frac{1}{2!}\frac{\partial ^{2}E(n_B
,\delta, n_s)}{\partial \delta_q ^{2}}\right\vert _{\delta_q =0}.
\label{QMEsym}
\end{equation}%
The isospin asymmetry of quark matter is defined as
\begin{equation}
\delta_q = 3\frac{n_d-n_u}{n_d+n_u},
\label{delta}
\end{equation}
which equals to $-n_3/n_B$ with the isospin density $n_3 = n_u-n_d$ and
$n_B = (n_u+n_d)/3$ for two-flavor $u$-$d$ quark matter. We note that the 
above definition of $\delta_q $ for quark matter has been extensively used 
in the literature~\cite{DiT06,Pag10,DiT10,Sha12,Chu14}, and one has
$\delta_q = 1$ ($-1$) for quark matter converted by pure neutron (proton) matter
according to the nucleon constituent quark structure, consistent with the
conventional definition for nuclear matter, namely, $\frac{\rho_n -\rho_p}{\rho_n +\rho_p}=-n_3/n_B$.
In Eq.~(\ref{EOSAQM}), the absence of odd-order terms in $\delta_q $ is due to
the exchange symmetry between $u$ and $d$ quarks in quark matter when
one neglects the Coulomb interaction among quarks. The higher-order
coefficients in $\delta_q $ are shown to be very small in various model 
calculations~\cite{Chu14}.

It has been demonstrated recently~\cite{Chu14} that the isovector properties of quark
matter may play an important role in understanding the properties of strange quark matter
and quark stars. If the recently discovered heavy pulsars PSR J1614-2230~\cite{Dem10} and 
PSR J0348+0432~\cite{Ant13} with mass around $2M_{\odot}$ were quark stars, they can put important 
constraint on the isovector properties of quark matter, especially the quark matter 
symmetry energy. Within the confined-isospin-density-dependent-mass
(CIDDM) model~\cite{Chu14}, in particular, it has been shown that the
two-flavor $u$-$d$ quark matter symmetry energy should be at least about twice that
of a free quark gas or normal quark matter within conventional NJL model in order to
describe the PSR J1614-2230 and PSR J0348+0432 as quark stars.

\section{Summary}

\label{summary}

We have explored the systematics of the density dependence of the
symmetry energy in the ambit of microscopic calculations with various 
energy density functionals. Our results indicate that the symmetry 
energy magnitude $E_{\text{sym}}({\rho })$ and its density slope 
$L({\rho })$ from subsaturation density to supra-saturation density 
can be essentially determined by three parameters defined at saturation 
density $\rho_0 $, i.e., the magnitude $E_{\text{sym}}({\rho_0 })$, 
the density slope $L$ and the density curvature $K_{\text{sym}}$. 
This finding implies that three values of $E_{\text{sym}}({\rho })$ 
or $L({\rho })$ essentially determine $E_{\text{sym}}({\rho })$
and $L({\rho })$ in large density region. In particular, using 
$E_{\text{sym}}({\rho_c}) = 26.65 \pm 0.2$ MeV
and $L({\rho_c}) = 46.0 \pm 4.5$ MeV at $\rho_{\rm{c}}= 0.11$ fm$^{-3}$ 
extracted from isotope binding energy difference and neutron skin of Sn
isotopes together with $E_{\text{sym}}({\rho_{0}}) = 32.5 \pm 0.5$ MeV 
obtained from finite-range droplet model analysis of nuclear binding energy, 
we obtain $L=46.7 \pm 12.8$ MeV and $K_{\text{sym}}=-166.9 \pm 168.3$ MeV 
as well as $E_{\text{sym}}({2\rho _{0}}) \approx 40.2 \pm 12.8$
MeV and $L({2\rho _{0}}) \approx 8.9 \pm 108.7$ MeV. These results favor a 
soft to roughly linear density dependence of the symmetry energy.

We have also discussed the quark matter symmetry energy, which has been 
shown to play an important role in understanding the properties of strange 
quark matter and quark stars. In particular, it has been suggested that the
two-flavor $u$-$d$ quark matter symmetry energy should be at least about 
twice that of a free quark gas or normal quark matter within conventional 
NJL model in order to describe the recently discovered heavy pulsars PSR 
J1614-2230 and PSR J0348+0432 with mass around $2M_{\odot}$ as quark stars.

The author thanks Wei-Zhou Jiang, Che Ming Ko, Bao-An Li, De-Hua Wen, 
Hermann Wolter, and Jun Xu for useful discussions. This work was supported 
in part by the Major State Basic Research Development Program (973
Program) in China under Contracts No. 2015CB856904 and
No. 2013CB834405, the National Natural Science Foundation
of China under Grants No. 11275125 and No. 11135011,
the ``Shu Guang'' project supported by Shanghai Municipal
Education Commission and Shanghai Education Development
Foundation, the Program for Professor of Special Appointment
(Eastern Scholar) at Shanghai Institutions of Higher Learning,
and the Science and Technology Commission of Shanghai
Municipality (11DZ2260700).

\end{document}